\documentclass[%
 reprint,
preprintnumbers,
 amsmath,amssymb,
 aps,prx, 
longbibliography, 
floatfix,
]{revtex4-1}

\usepackage{graphicx}
\usepackage{epstopdf}
\usepackage{dcolumn}
\usepackage{bm}
\usepackage{hyperref}
\usepackage[normalem]{ulem}

\newcommand{\ie}{\textit{i.e.}, }
\newcommand{\eg}{\textit{e.g.}, }

\newcommand{\etal}{\textit{et al.} }
\newcommand{\tin}{\mathrm{in} }
\newcommand{\tout}{\mathrm{out} }

\begin{document}


\title{Plasma $q$-plate for generation and manipulation of intense optical vortices}
\author{Kenan Qu}
\author{Qing Jia}%
\author{Nathaniel J. Fisch}%
\affiliation{Department of Astrophysical Sciences, Princeton University, Princeton, New Jersey 08544, USA }%

\date{\today}

\begin{abstract}
	An optical vortex is a light wave with a twisting wavefront around its propagation axis and null intensity in the beam center. 
	Its unique spatial structure of field lends itself to a broad range of applications, including optical communication, quantum information, superresolution microscopy, and multi-dimensional manipulation of particles. However, accessible intensity of optical vortices have been limited to material ionization threshold. This limitation might be removed by using the plasma medium. 
	Here we propose the design of suitably magnetized plasmas which, functioning as a $q$-plate, leads to a direct convertion from a high-intensity Gaussian beam into a twisted beam. A circularly polarized laser beam in the plasma accumulates an azimuthal-angle-dependent phase shift and hence forms a twisting wavefront. Our three-dimensional particle-in-cell simulations demonstrate extremely high power conversion efficiency. The plasma $q$-plate can work in a large range of frequencies spanning from terahertz to the optical domain. 
\end{abstract}

\maketitle


\section{Introduction}

Manipulating intense laser beam using plasma as a medium has unparalleled advantages compared with conventional solid-state media because plasma can sustain ultra-high intensities. Impressive applications in plasmas include backward Raman amplification~\cite{PRL-Shvets1998, PRL-Malkin1999, Ren_08, ExpPRL2004-SRA, ExpPRL2004-Ping, ExpPRL2005-Suckewer, ExpNat2007-Suckewer, EXPPRL2008-Pai}, X-ray lasers~\cite{Xray-wang-2008, Xray-Vartanyants-2011, Xray-Singer-2012}, plasma gratings~\cite{PlasmaGratings2014, Lehmann2016, Lehmann2017}, and plasma holography~\cite{PRL2001-Dodin}. Recently, there has been increasing interest in generating and manipulating optical vortices~\cite{nye1974OV, coullet1989OV, gahagan1996OV, allen1992OAM}, which carry orbital angular momentum (OAM), pointing to a variety of applications, including trapping~\cite{gahagan1996OV} and rotating~\cite{coullet1989OV, madison2000vortex} suitable materials particles, imaging and probing physical and biological properties of matters~\cite{curtis2002tweezers, foo2005coronagraph, dholakia2011shaping, wang2011bio}, improving communication bandwidth~\cite{yan2014multiplexing}, and encoding quantum information in higher-dimensional Hilbert spaces~\cite{gibson2004quantum, groblacher2006quantum}.

Manipulation of laser wavefronts for generating optical vortcies requires creating structured optical anisotropicity. The task is more challenging in plasma because the medium is inherently unstructured~\cite{photonics4020028, SciRep_Qin, PPCF_Mendonca}. Methods of producing intense optical vortices in plasma~\cite{PRL_Shi, PRL_Zhang, Leblanc2017} had relied exclusively on a plasma mirror structure, which quickly deforms due to plasma expansion. Vieira~\etal~\cite{NatC_Vieira, PRL_Mendonca, PRL_Vieira, Mendonca_PoP2015} proposed that the use of stimulated Raman scattering in plasmas can lead to the amplification of optical vortices to very high powers and OAM charge numbers. However, the power conversion efficiency is restricted by wave-wave coupling efficiency and subject to plasma instabilities~\cite{Kenan_plasmaSeed}. What is needed to process high-power OAM is to employ plasma, but without reliance upon the nonlinear interaction of plasma waves.

Actually, the azimuthal anisotropicity required for OAM can be provided by the laser beam itself: A circularly polarized light has an intrinsic twisting phase structure which carries spin angular momentum (SAM). 
Remarkably, there exists a phase-only optical element, called the $q$-plate~\cite{PRL_qplate}, which can convert SAM of light into OAM:
\begin{equation}
	E_0(r)\mathbf{\hat{e}}_{L,R} \xrightarrow{\text{$q$-plate}} E_0(r)\exp(il\varphi) \mathbf{\hat{e}}_{R,L}, 
\end{equation} 
where $r,\varphi$ are the polar coordinates in the $xy$-plane, and $\mathbf{\hat{e}}_{L,R}$ is the unit vector for left-hand (LH) and right-hand (RH) circular polarizations.  The OAM helicity number  (also called charge number) is denoted by $l$, and the SAM helicity number is $\sigma = \pm 1$ for LH/RH circular polarizations. A $q$-plate is a thin optical birefringent  phase plate with its fast axis perpendicular to the laser propagation direction. Its fast axes have certain topological structures, rather than a homogeneous structure like a half-wave plate. A circularly polarized laser beam after passing through the $q$-plate is converted to one with the opposite circular polarization and, more importantly, to a helical wavefront. 

We propose a plasma $q$-plate where a magnetic field controls the optical fast axes. We demonstrate numerically that it can generate an optical vortex with OAM from a circularly polarized Gaussian laser. 
Such a plasma $q$-plate has an intensity limit set by relativistic rather than ionization effects, so that an ultra-intense OAM laser beam can be generated. The laser mode conversion relies on the anisotropicity of the dispersion relation in a magnetic field, but does not require a resonant wave-wave interaction. This avoids preparing exact wavevector and frequency matching for lasers and plasmas, as required in previous plasma-based generation schemes~\cite{PRL_Shi, PRL_Zhang, NatC_Vieira, PRL_Mendonca, PRL_Vieira, Leblanc2017}, thereby reducing the experimental complexity and improving the engineering flexibility.

\begin{figure*}[htbp]
	\centering
	\includegraphics[width=0.95\textwidth]{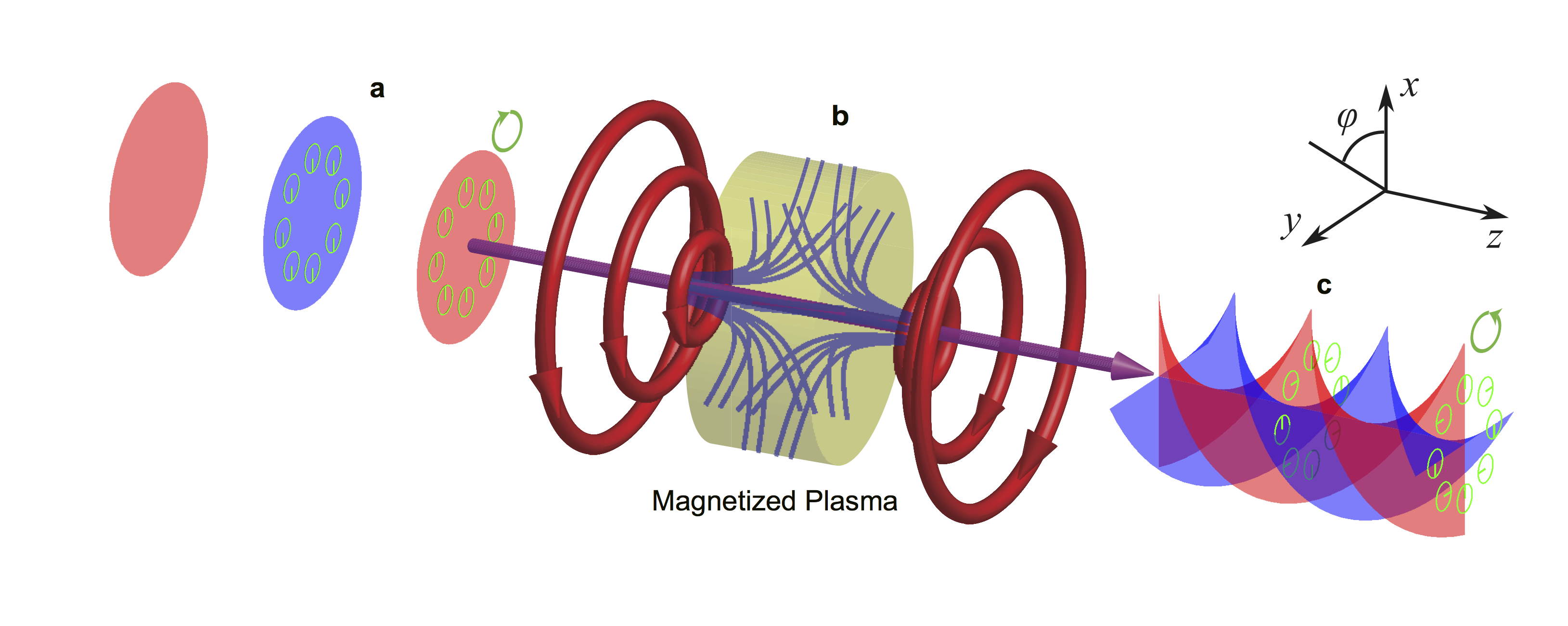}
	\caption{\label{fig1} Converting a Gaussian laser beam into an LG beam in magnetized plasma. An input Gaussian laser beam ({\bf a}) is sent through a plasma, which is mediated in an axial symmetric magnetic field generated by anti-Helmholtz coils ({\bf b}). The wavefront of the output laser beam (shown in {\bf c})  becomes twisted. In {\bf a} and {\bf c}, the light red and blue shades show the isosurfaces of the wavefront in which the electric fields are parallel and perpendicular to the azimuthal directions, respectively. The small green circles show the polarization. The ticks show the instantaneous directions of the electric fields calculated using Eq.~(\ref{eqeout}). In \textbf{b}, the green shaded cylinder is the plasma and grey lines illustrate the magnetic field lines. }
\end{figure*}

\section{Scheme}

The proposed plasma $q$-plate is an optical element which manipulates the laser phase through the optical birefringence that is induced by the external magnetic field in plasmas. The birefringence arises from different dispersion relations of two eigenmodes of a laser when it propagates perpendicular to the external magnetic field $\mathbf{B}_0$ into a plasma. When the laser polarization is parallel to $\mathbf{B}_0$, its propagation is not affected by the magnetic field and its refractive index is $n_\| = \sqrt{1-\omega_p^2/\omega^2}$. Here, $\omega$ is the laser frequency, $\omega_p=\sqrt{n_e e^2/m_e\varepsilon_0}$ is the plasma frequency where  $e$ is the natural charge, $c$ is the speed of light, $\varepsilon_0$ is the vacuum permittivity, $n_e$ and $m_e$ are the electron density and mass, respectively. 
When the laser polarization is perpendicular to $\mathbf{B}_0$, the gyromotion of electrons hybridizes the electromagnetic mode and electrostatic mode, and the refractive index becomes $n_\perp = \sqrt{1- \frac{\omega^2-\omega_p^2}{\omega^2-\omega_c^2-\omega_p^2} \bigg(\frac{\omega_p^2}{\omega^2}\bigg)} \displaystyle$. Here, $\omega_c = eB_0/m_e c$ is the electron gyrofrequency. 
The difference of refractive indices for different polarization eigenmodes, 
\begin{equation}\label{1}
\Delta n=n_\|-n_\perp, 
\end{equation}
induces spatially varied phase shifts to the laser wavefront depending on the angle between laser polarizations and magnetic field directions. The end result is optical birefringence, which converts a circularly polarized laser beam into a linearly polarized one and further into a circularly polarized one with the opposite polarization chirality. The magnetic field lines constitute the slow axes of birefringence.

To create a twisting laser wavefront, the plasma $q$-plate needs to impose an azimuthally varying phase shift to the Gaussian beam or plane wave.
Specifically, we consider an axial symmetric magnetic field whose lines are along the azimuthal directions $\hat{\varphi}$. A convenient way to produce the required anisotropic magnetic field is to use anti-Helmholtz coil pairs, as illustrates in Fig.~\ref{fig1}. Each anti-Helmholtz coil pair consists of two parallel coils carrying currents in the opposite directions. The magnetic field in the middle of the coils is purely radial. 
By carefully arranging multiple pairs of anti-Helmholtz coils with different radii and currents, one can produce an equal-amplitude magnetic field within a certain range of radii (see {\bf Appendix A} for further details). The profile of the magnetic field in our study is similar to a CUSP geometry~\cite{CUSP_1957, CUSP_1959}, which ensures magnetohydrodynamic stability of plasma.

\begin{figure*}[htb]
	\centering
	\includegraphics[width=0.99\textwidth]{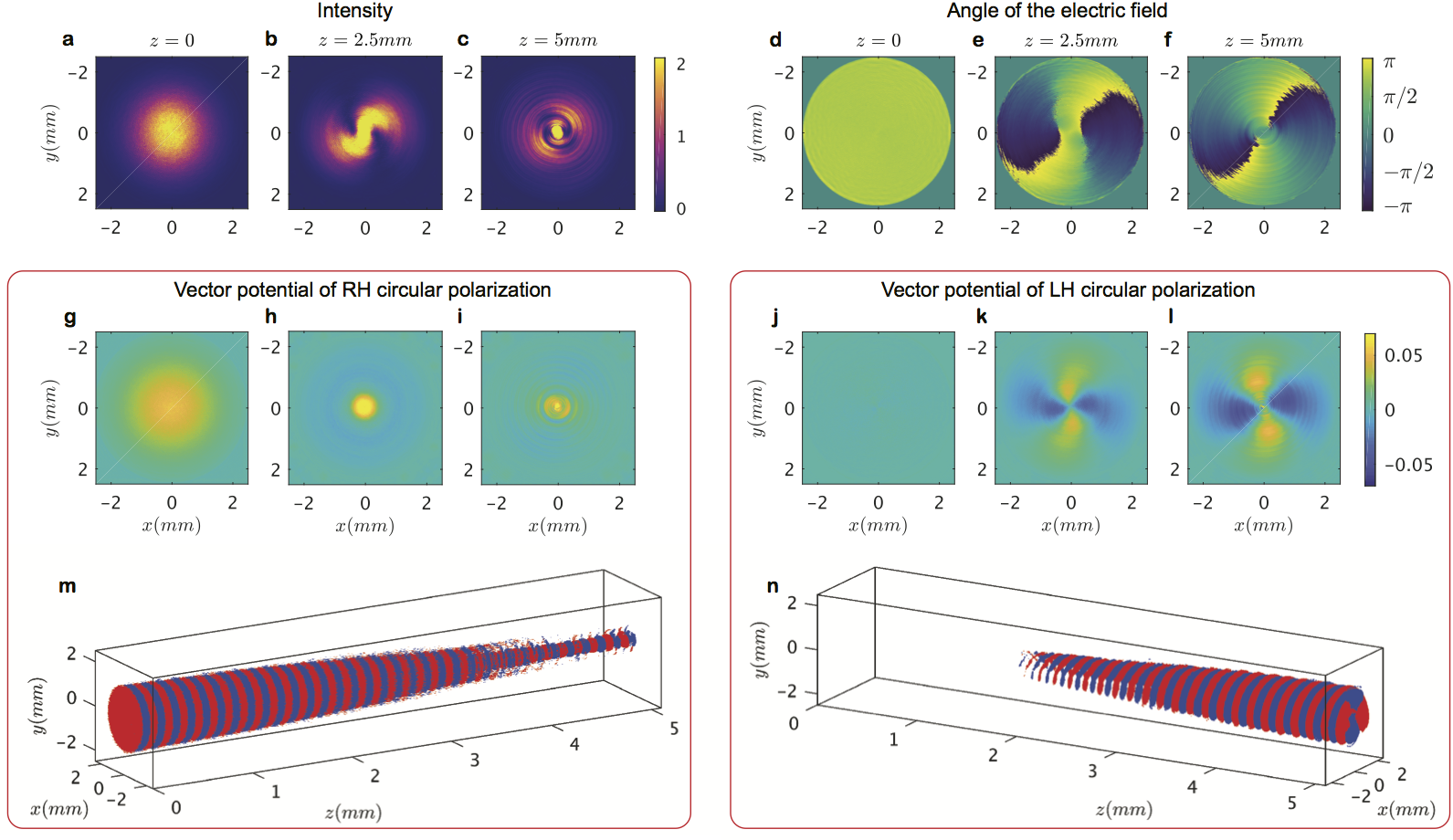}
	\caption{\label{fig2} PIC simulation results of converting a Gaussian laser beam into an optical vortex using a plasma $q$-plate. \textbf{a}-\textbf{c} and \textbf{d}-\textbf{f} show the beam intensity profile and angle of the electric field with respect to the x-axis at different cross-sections (labelled on top of each plot), respectively. \textbf{g}-\textbf{i} and \textbf{j}-\textbf{l} show the normalized vector potential $a$'s of the RH and LH circularly polarized components at different cross-sections, respectively. The three-dimensional plots \textbf{m} and \textbf{n} show isosurface of the normalized vector potential field with red and blue colors denote $a=\pm 0.03$, respectively. The input beam (\textbf{m}) has a circular polarization and beam waist of $5\,\mu$m. The output beam (\textbf{n}) shows a double spiral structure.  }
\end{figure*}

\begin{figure*}[hbt]
	\centering
	\includegraphics[width=0.99\textwidth]{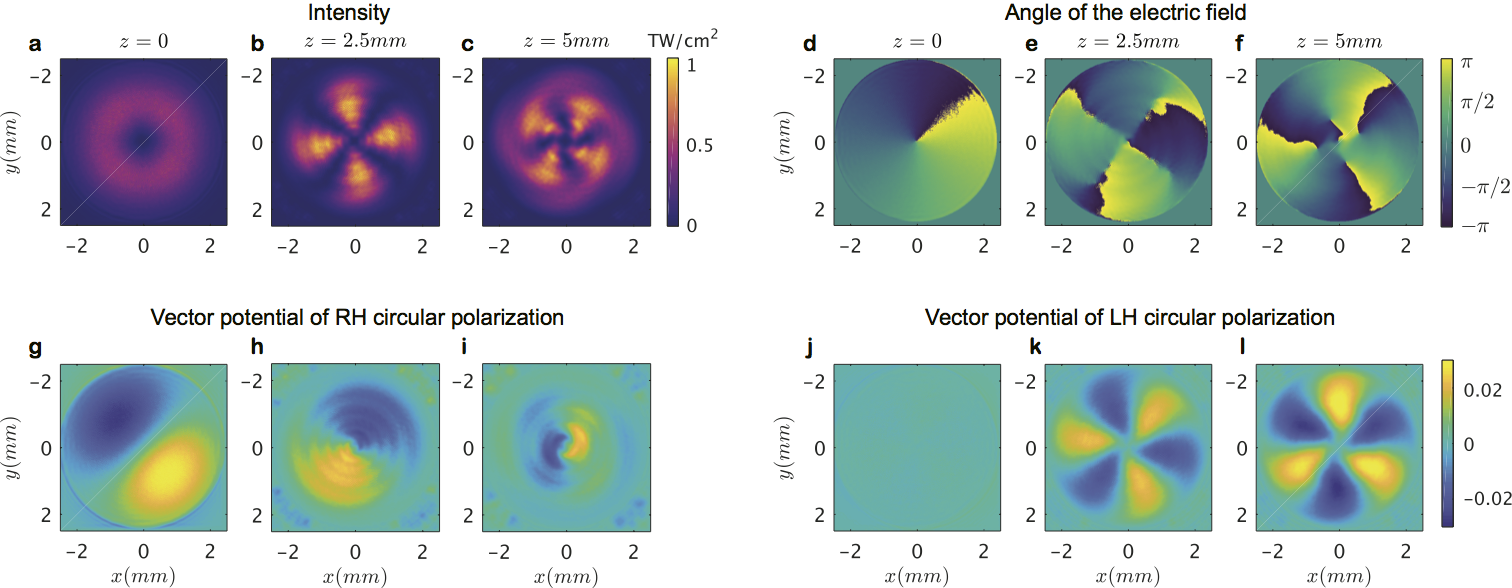}
	\caption{\label{fig3} PIC simulation results of converting the charge number of an LG laser beam from $l=-1$ to $l=-3$ using a plasma $q$-plate. \textbf{a}-\textbf{c} and \textbf{d}-\textbf{f} show the beam intensity profile and angle of the electric field with respect to the x-axis at different cross-sections (labelled on top of each plot), respectively. \textbf{g}-\textbf{i} and \textbf{j}-\textbf{l} show the normalized vector potential $a$'s of the RH and LH circularly polarized components at different cross-sections, respectively. The parameters for the magnetized plasma and laser beam are identical to Fig.~\ref{fig2} except the initial OAM charge number.  }
\end{figure*}

The input laser for the $q$-plate is a circularly polarized Gaussian beam which has a homogeneous phase front. Without losing generality, we take the example of an LH circular polarization. In Fig.~\ref{fig1}, we use red and blue shaded planes to denote the phase isosurfaces, in which the electric field oscillates in the same directions shown as green ticks in the polarization circles. Locally, the electric field at each azimuthal angle can be decomposed as the superposition of an ordinary polarization along the direction of $\mathbf{B}_0$ and an extraordinary polarization perpendicular to $\mathbf{B}_0$. Note that the external magnetic field is anisotropic hence the decomposition of laser polarization depends on the azimuthal angle. Each phase isosurface plane is divided into four quadratures which are consecutively dominated by ordinary polarization and extraordinary polarization. In plasma, the extraordinary polarization propagates at a larger phase velocity and hence splits the laser isosurface leading to a curved wavefront. 
At the output, the extraordinary polarization accumulates a $\pi$-phase shift and flips its direction. Since the ordinary polarization maintains its direction, the output beam polarized becomes RH circularly polarized. The wavefront curvature becomes so large that two adjacent phase isosurfaces of the same ``color'' connect to each other thereby creating a continuous helical wavefront, as shown in Fig.~\ref{fig1}{\bf c}.

More rigorous analysis of the optical effect of a magnetized plasma can be carried out using the Jones formalism~\cite{hecht2002optics}. It can straightforwardly describe the evolution of instantaneous electric field of fully polarized light when it passes through optical devices. An input laser $E_\tin$, which takes an LH circularly polarized profile, can be represented by a Jones vector  $\displaystyle E_\tin = \binom{E_x}{E_y} = \binom{1}{i} E_0$ where $E_0$ is the electric field of the input beam. The two vector components $E_x$ and $E_y$ denote its horizontal and vertical polarization with equal amplitude and a $\pi/2$-phase difference. 
The plasma channel of an arbitrary thickness, in general, partially transforms an LH circular polarization into an RH one, or vice versa. Its optical property can be modeled by a phase retarder
\begin{align}
M &= (\cos\frac{\xi}{2}) I + (\sin\frac{\xi}{2}) H, \\
I &= \begin{pmatrix} 1 & 0 \\ 0 & 1 \end{pmatrix}, \qquad
H = \begin{pmatrix}
\cos2\varphi & \sin2\varphi \\ \sin2\varphi & -\cos2\varphi
\end{pmatrix}, \nonumber
\end{align}
where $H$ is the Jones matrix of a half-wave plate with its fast axis at the angle $\varphi$, and $\xi = 2\pi L\Delta n/\lambda$ is the phase retardation between the fast and slow axes. $L$ and $\lambda$ denote the plasma length and laser wavelength in vacuum, respectively. The output beam is determined by the matrix product 
\begin{equation}\label{eqeout}
E_\tout = M E_\tin = (\cos\frac{\xi}{2}) \binom{1}{i} E_0 + (\sin\frac{\xi}{2}) e^{i2\varphi} \binom{1}{-i} E_0. 
\end{equation}
If $\xi$ is an odd integer multiple of $\pi$, the $\cos(\xi/2)$ term vanishes indicating that the LH polarized input beam is fully converted into a mode with the opposite polarization. More importantly, the output mode reveals an extra $\varphi$-dependent phase shift which up-converts the topological charge number by $2$. The result in Eq.~(\ref{eqeout}) also shows that the input and output OAM modes have the opposite circular polarizations. This property allows one to isolate the generated optical vortex from the input Gaussian beam by using, \eg a polarized beam-splitter.

\section{Three-dimensional simulations}

To demonstrate the wavefront manipulation using a plasma $q$-plate, we conduct particle-in-cell (PIC) simulations of laser plasma interaction using code EPOCH~\cite{EPOCH2015}. The input laser beam is a high-intensity ($2.74\, \mathrm{TW/cm}^2$) RH circularly polarized Terahertz Gaussian beam. It has a frequency of $\omega=2\pi\times 3$ THz and wavelength of $\lambda=100\,\mu\mathrm{m}$. 
It is focused at the center of the plasma channel, with a waist of $1.6$ mm at the focal plane.  
The plasma channel has an electron density $n_e=6.2779\times10^{16}\, \mathrm{cm}^{-3}$ and plasma frequency $\omega_p = 0.75 \omega$. 
The external magnetic field takes an axial symmetric structure which is generated by three anti-Helmholtz coil pairs (see {\bf Appendix B} for more details). Specifically, its radial component has magnitude $B_r=10$ T and the corresponding cyclotron frequency is $\omega_c=0.1\omega$. Its axial component is nonzero only at $r<50\mu$m and $B_z = 20(x/\mu\mathrm{m} -50)$ T. 
With this set of parameters, the birefringence is $\Delta n=0.01$, which determines the plasma length $L=\lambda/(2\Delta n)=5$ mm.

We depict in Fig.~\ref{fig2}\textbf{a}-\textbf{c} and \textbf{d}-\textbf{f} a snapshot of the optical intensity and angle of the electric fields at the time $t=36$ ps when the laser front exits the plasma. 
Figure~\ref{fig2}\textbf{a} shows the input Gaussian beam ($l=0$) which has an axial symmetric transverse profile with an intensity peak located in the beam center. When it enters plasma, the Gaussian mode begins to transform into a Laguerre-Gaussian (LG) mode which has an intensity null in the center and peak in the periphery. In the middle of the plasma channel where the conversion ratio is near $50$\%, the two different modes superimpose creating a narrow line-shape intensity peak. In the same regions, the direction of electric field observed from Fig.~\ref{fig2}\textbf{e} is a constant revealing linear polarization. 
The intensity profile at the output cross-section is shown in Fig.~\ref{fig2}\textbf{c}. We can clearly see that the intensity peak forms a ring of radius of about $0.7$ mm. The electric fields oscillate only at tangential angles, as shown in Fig.~\ref{fig2}\textbf{f}. Thus, we conclude that the output beam has a large component with an LG mode with $l=2$. The power conversion efficiency is found to be as high as $83$\%. 
We also note that the output beam has a finite intensity in the beam center within the radius of about $0.2$ mm, as shown in Fig.~\ref{fig2}\textbf{c}. The small spot of intensity peak is the residual of non-converted input beam. This nonconversion is evident also from Fig.~\ref{fig2}\textbf{f}.

Wavefront manipulation using a $q$-plate is accompanied by the change of different circular polarizations. We evaluate the mode conversion by decomposing the laser beam into the LH and RH circular polarizations and depict each component in Fig.~\ref{fig2}. The decomposition is calculated by adding up the electric fields of two planes separated at a quarter wavelength distance, \ie $[E_\varphi(z) \pm E_{\varphi-\pi/2} (z-\lambda'/4)] /\sqrt{2}$ where $\lambda'=150\,\mu$m is the laser wavelength in plasma and $\pm$ determines the helicity of polarization. The OAM of light can be witnessed from its instantaneous electric fields, \eg $E_x(\mathbf{r})$. It is usually measured from the interference pattern of itself and a linearly polarized reference beam at the same frequency. The wavefronts of each circularly polarized component are shown in Fig.~\ref{fig2}\textbf{g}-\textbf{n} using their vector potentials normalized as $a=\lambda(\mu\mathrm{m}) \sqrt{I(\mathrm{W/cm}^2)/2.74 \times10^{18}}$. The sign of $a$ is determined by the local electric field of the corresponding polarization component. It is evident that RH circular polarization is associated with the Gaussian beam profile and it is transformed into the LG mode only when converted into the LH circular polarization. The mode conversion immediately begins when the beam enters plasma. The LH circular polarized component in the output, shown in Fig.~\ref{fig2}\textbf{l}, is characterized by a four-lobe structure and two adjacent lobes have the opposite phases.

Figures~\ref{fig2}\textbf{m} and \textbf{n} show three-dimensionally the iso-amplitude surface of the RH and LH circular polarization components. The mode conversion can be clearly seen from the changes of their spot sizes along the propagation direction. Note that the mode conversion saturates at about $z=4$ mm. The central region retains its intrinsic and structural helicity, while mode conversion continues in the peripheral regions. This is because,  near the central axis,  the dispersion relation significantly deviates from the idealized magnetic geometry assumed in Eq.~(\ref{1}).

A unique property of the specific design of plasma $q$-plate is that each photon does not transfer any optical torque into plasma, although both of its SAM and OAM are converted by the plasma. Actually, its SAM is exactly converted into OAM which ensures the conservation of total angular momentum. Continued propagation of the output OAM beam in the magnetized plasma will convert it into the original RH polarization and $l=0$. However, if the input beam has both SAM and OAM with identical sign of helicity, a higher-order LG output beam can be generated with $l$ decreased by $2$. 

Figure~\ref{fig3} shows higher-order OAM generation. The simulation set-up follows the example of Fig.~\ref{fig2}, except the input laser is replaced with one of $l=-1$. The beam has a ring-shape intensity profile (shown in Fig.~\ref{fig3}\textbf{a}) and a single helical plane of electric fields (shown in Fig.~\ref{fig3}\textbf{d}). Decomposing the input field into two opposite circularly polarizations yields only the RH polarized component, shown in Fig.~\ref{fig3}\textbf{g}. During propagation, it can be observed that the RH component is gradually converted into LH polarized. At the same time, the two-lobe structure (shown in Fig.~\ref{fig3}\textbf{h}-\textbf{i}) transfers into a six-lobe profile (shown in Fig.~\ref{fig3}\textbf{k}-\textbf{l}). The phase plot of electric fields angles eventually shows three singularities spinning in the RH direction, indicating that the output beam has a charge number $l=-3$.

\begin{figure}[t]
	\centering
	\includegraphics[width=0.5\textwidth]{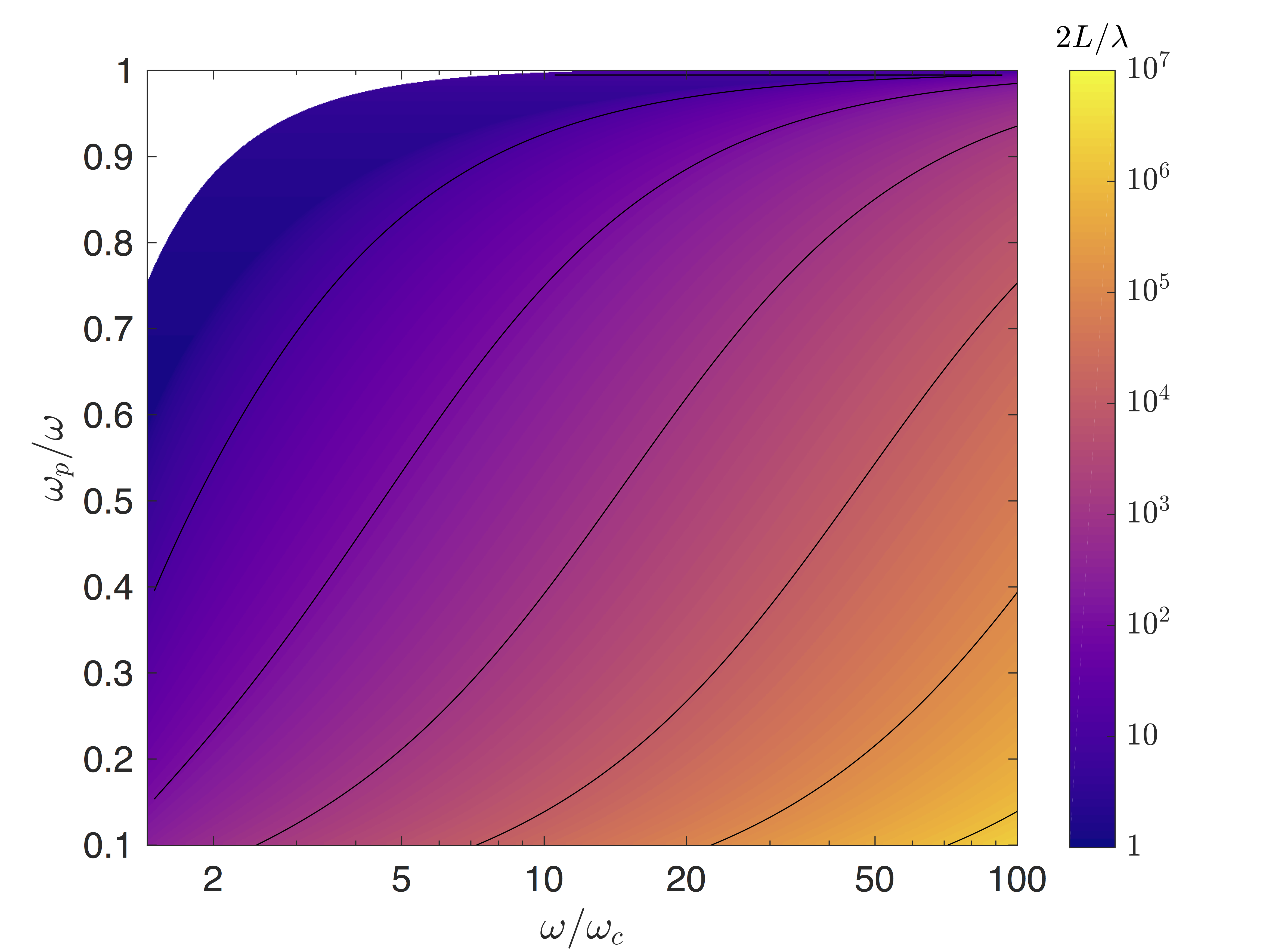}
	\caption{\label{fig4} Contour plot of the plasma length for complete conversion with different plasma frequencies $\omega_p$ and gyrofrequencies $\omega_c$ normalized to the laser frequency $\omega$. The laser does not propagate in the blank region on the top left corner.  }
\end{figure}

\section{Conclusions}

While the mechanism is only simulated here, we can anticipate that the plasma $q$-plate will operate over a large range of laser frequencies, spanning from terahertz to infrared optical frequencies.  The large range is enabled by the multiplicity of free parameters to choose, namely, plasma density, plasma length, and magnetic field strength. For a complete laser mode conversion, the optimal plasma length is $L= \lambda/(2\Delta n)$. Its dependence on the laser frequency $\omega$, plasma frequency $\omega_p$ and cyclotron frequency $\omega_c$ is shown in Fig.~\ref{fig4}. We find that  the optimal plasma length $L$ decreases with larger $\omega_c$ and $\omega_p$ until the laser frequency $\omega$ is below the ``cut-off'' frequency $(\omega_c + \sqrt{\omega_c^2 + 4\omega_p^2})/2$. Practically, the frequency of a homogeneous plasma can be controlled below $0.75\omega$ and its length below $10^5\lambda$. For fixed external magnetic field amplitude, the operational frequency for the laser is hence between $\omega_c$ to $100\omega_c$. Considering that the magnetic fields generated by anti-Helmholtz coils are limited to a few Tesla (sub-THz cyclotron frequency), the range of laser frequency can be up to a few hundred THz. For example, the $q$-plate with the same plasma and magnetic field as used in Fig.~\ref{fig2} also works for a $1\,\mu$m laser with frequency $\omega= 2\pi\times 300$ THz. 

Note that the $q$-plate mechanism is solely based on the difference of optical path lengths for ordinary and extraordinary polarizations. Given a uniform magnetic field within the path of laser beam, plasma density fluctuation leads to variance of refractive index and birefringence. However, the condition of density homogeneity in the longitudinal direction is not important since the power conversion ratio is proportional to $\cos(\xi)$ according to Eq.~(\ref{eqeout}), which suppresses the fluctuation of $\xi$ to the second order. Imperfect conversion due to large density fluctuation can be compensated by adjusting the strength of the magnetic field. A more relevant concern is the restriction on the transverse homogeneity of plasma density; it leads to a mixture of high-order modes reducing the beam quality. In practice, plasma channels often have higher density in the center than its periphery. A possible solution is to appropriately arrange the strength of magnetic field with lower magnitude in the center in order to retain homogeneous birefringence.

\section*{Appendix A: Anti-Helmholtz Coils}\label{appA}

An anti-Helmholtz coil (AHC) pair is formed by two identical ring-shape wires shifted vertically and carrying current in opposite directions. The magnetic field for each circular current loop of current $I$ and radius $R$ displaced from the center of two loops by a distance $D$ is given by 
\begin{align}
	B_z =& \frac{\mu I}{2\pi} \frac{1}{\sqrt{(R+\rho)^2 + (z-D)^2}} \nonumber \\
	& \qquad \times \left[ K(k^2) \frac{R^2-\rho^2 -(z-D)^2}{(R-\rho)^2 +(z-D)^2} E(k^2) \right], \\
	B_\rho =& \frac{\mu I}{2\pi} \frac{1}{\rho} \frac{z-D}{\sqrt{(R+\rho)^2 + (z-D)^2}} \nonumber \\
	& \qquad \times \left[ -K(k^2) + \frac{R^2+\rho^2 +(z-D)^2}{(R-\rho)^2 +(z-D)^2} E(k^2) \right],
\end{align}
where 
\begin{equation}
k^2 = \frac{4R\rho}{(R+\rho)^2 + (z-D)^2},
\end{equation}
$K(k^2)$ and $E(k^2)$ are the complete elliptic integrals for the first and second kind, respectively, $\rho$ is the axial distance and $z$ is the height. 
The $B_\rho$ near the middle of the AHC increases at $\rho<R$ and decreases when $\rho>R$. The peak at $\rho=R$ allows a quasi-homogeneous $B_\rho$ in a range of radii by combining multiple AHC pairs in parallel. For example, three AHC pairs with radii and currents ($0.3R$, $I/50$), ($0.5R$, $I/15$), and ($R$, $I$), respectively, can create a constant $B_\rho$ in the range $0.3R$ to $1.3R$ with a fluctuation below $10\%$. The $B_z$ component has a linear gradient in the axial direction and is zero in the plane in the middle of the AHC pairs.

\section*{Appendix B: Particle-in-cell simulations} \label{appB}

The three-dimensional (3D) particle-in-cell (PIC) simulations are conducted using the kinetic, full-relativistic codes EPOCH. The simulation box dimensions are $5\,\mathrm{mm}\times 5\,\mathrm{mm}\times 6\,\mathrm{mm}$ in the $x-y-z$ directions and it has been divided into $8000\times 8000\times 9600$ cells with a cell size of $1/16$ of the laser wavelength in each dimension. The electrons are distributed in a cylinder in the center of the simulation box along the $z$ direction,  with a radius of $2.45\,\mathrm{mm}$ in the $x-y$ plane and a length of $5.5\,\mathrm{mm}$ in the $z$ direction. Each cell in the cylinder contains $20$ electrons and $20$ protons. The space outside the cylinder is set as vacuum. The electron temperature is $10\, \mathrm{eV}$ and proton temperature is $0.1\, \mathrm{eV}$.

\begin{acknowledgments}
This work was supported by NNSA Grant No. DE-NA0002948, and AFOSR Grant No. FA9550-15-1-0391. Computing support for this work came from the Lawrence Livermore National Laboratory (LLNL) Institutional Grand Challenge program. The EPOCH  code was developed as part of the UK EPSRC grants EP/G054950/1,
EP/G056803/1, EP/G055165/1 and EP/ M022463/1.
\end{acknowledgments}

\bibliography{oam}

\end{document}